\documentclass[prl,twocolumn,showpacs,floatfix,longbibliography]{revtex4-1}
\usepackage{mathrsfs}
\usepackage{amssymb, amsbsy, amsmath, latexsym, dsfont, array, layout,mathrsfs,color,ulem,bm}
\usepackage[colorlinks,linkcolor=blue,anchorcolor=red,citecolor=blue]{hyperref}

\newcommand{\one}{\mathds{1}}
\newcommand{\ket}[1]{\left|{#1}\right\rangle}
\newcommand{\bra}[1]{\left\langle{#1}\right|}

\newcommand{\ketbrad}[1]{\left|{#1}\rangle\!\langle{#1}\right|}

\usepackage{graphicx}
\usepackage{xspace, stmaryrd, ulem}
\definecolor{delete}{rgb}{1.0, 0.0, 0.0}
\definecolor{edit}{rgb}{0.0, 0.0, 0.9}
\definecolor{comment}{rgb}{0.9, 0.0, 0.0}

\begin{document}

\title{Detecting Non-Bloch Topological Invariants in Quantum Dynamics}

\author{Kunkun Wang\textsuperscript{1,4}}\thanks{These authors contributed equally to this work}
\author{Tianyu Li\textsuperscript{2,3}}\thanks{These authors contributed equally to this work}
\author{Lei Xiao\textsuperscript{1}}
\author{Yiwen Han\textsuperscript{2,3}}
\author{Wei Yi\textsuperscript{2,3}}\email{wyiz@ustc.edu.cn}
\author{Peng Xue\textsuperscript{1}}\email{gnep.eux@gmail.com}

\affiliation{\textsuperscript{1}Beijing Computational Science Research Center, Beijing 100084, China}
\affiliation{\textsuperscript{2}CAS Key Laboratory of Quantum Information, University of Science and Technology of China, Hefei 230026, China}
\affiliation{\textsuperscript{3}CAS Center For Excellence in Quantum Information and Quantum Physics, Hefei 230026, China}
\affiliation{\textsuperscript{4}School of Physics and Optoelectronics Engineering, Anhui University, Hefei 230601, China}

\begin{abstract}
Non-Bloch topological invariants preserve the bulk-boundary correspondence in non-Hermitian topological systems, and are a key concept in the contemporary study of non-Hermitian topology. Here we report the dynamic detection of non-Bloch topological invariants in single-photon quantum walks, revealed through the biorthogonal chiral displacement, and crosschecked with the dynamic spin textures in the generalized quasimomentum-time domain following a quantum quench. Both detection schemes are robust against symmetry-preserving disorders, and yield consistent results with theoretical predictions. Our experiments are performed far away from any boundaries, and therefore underline non-Bloch topological invariants as intrinsic properties of the system that persist in the thermodynamic limit. Our work sheds new light on the experimental investigation of non-Hermitian topology.
\end{abstract}

\maketitle

{\it Introduction.---}Topology in open quantum systems acquires fascinating features that are absent in closed systems. A particularly intriguing example
is the breakdown of bulk-boundary correspondence---a fundamental principle for conventional topological matter~\cite{kane,Qi,OP}---in generic non-Hermitian topological systems. Therein, under the non-Hermitian skin effect, all nominal bulk states become localized toward boundaries~\cite{WZ1,Budich,alvarez,mcdonald,ThomalePRB}, rendering the conventional Bloch topological invariants that are defined on the Brillouin zone (BZ) ineffective in predicting the presence of edge states~\cite{Lee,WZ2,murakami,XZ,XR,fangchenskin,kawabataskin,tianshu,YZF2020,KSU19}.
To restore the bulk-boundary correspondence in these non-Hermitian systems, a non-Bloch band theory is introduced, prescribing non-Bloch topological invariants on a generalized Brillouin zone (GBZ) that takes into account the deformation of bulk states under open boundaries~\cite{WZ1,WZ2,murakami,tianshu,YZF2020}.
While both non-Hermitian skin effect and bulk-boundary correspondence have been observed in recent experiments~\cite{teskin,teskin2d,metaskin,photonskin,scienceskin},
the question remains whether non-Bloch topological invariants can be directly probed.
Such a question is motivated by existing experiments in closed systems, where topological invariants and phase transitions are dynamically probed through quantities such as statistical moments~\cite{statis}, mean chiral displacements~\cite{sciencecd,nccd,YB}, and emergent topological structures following a quantum quench~\cite{Weitenberg17,Weitenberg18,chenshuai,XPdqpt,XWH20,JZW20,xiongjun21}.

In this work, we report the experimental detection of non-Bloch topological invariants through dynamic signatures in single-photon quantum walks. We adopt two independent detection schemes, focusing respectively on a biorthogonal extension of the mean chiral displacement, and the dynamic spin textures in the GBZ following a quantum quench. While the former derives from the chiral symmetry of the quantum walk~\cite{ZCW}, the latter is characterized by a non-Bloch quantum quench theory~\cite{prr}.
The measured non-Bloch topological invariants from either scheme agree well with those predicted using the non-Bloch band theory. Remarkably, both schemes rely on the reality of the quasienergy spectra which are only possible under an open boundary condition. Our experiment is thus closely related to the recently established non-Bloch parity-time (PT) symmetry~\cite{nonBPT1,nonBPT2,songfei,EP}, where the exact PT phase and PT transitions are only allowed in the presence of boundaries, and are characterized by the non-Bloch band theory.
Indeed, we find that the measured dynamic spin textures in the GBZ further reveal the location of the non-Bloch exceptional points that separate the exact and broken PT phases.
Our results therefore confirm the non-Bloch topological invariant as an observable quantity in the bulk, and exemplify the applicability of non-Bloch band theory in dynamic processes.

\begin{figure}
\includegraphics[width=0.45\textwidth]{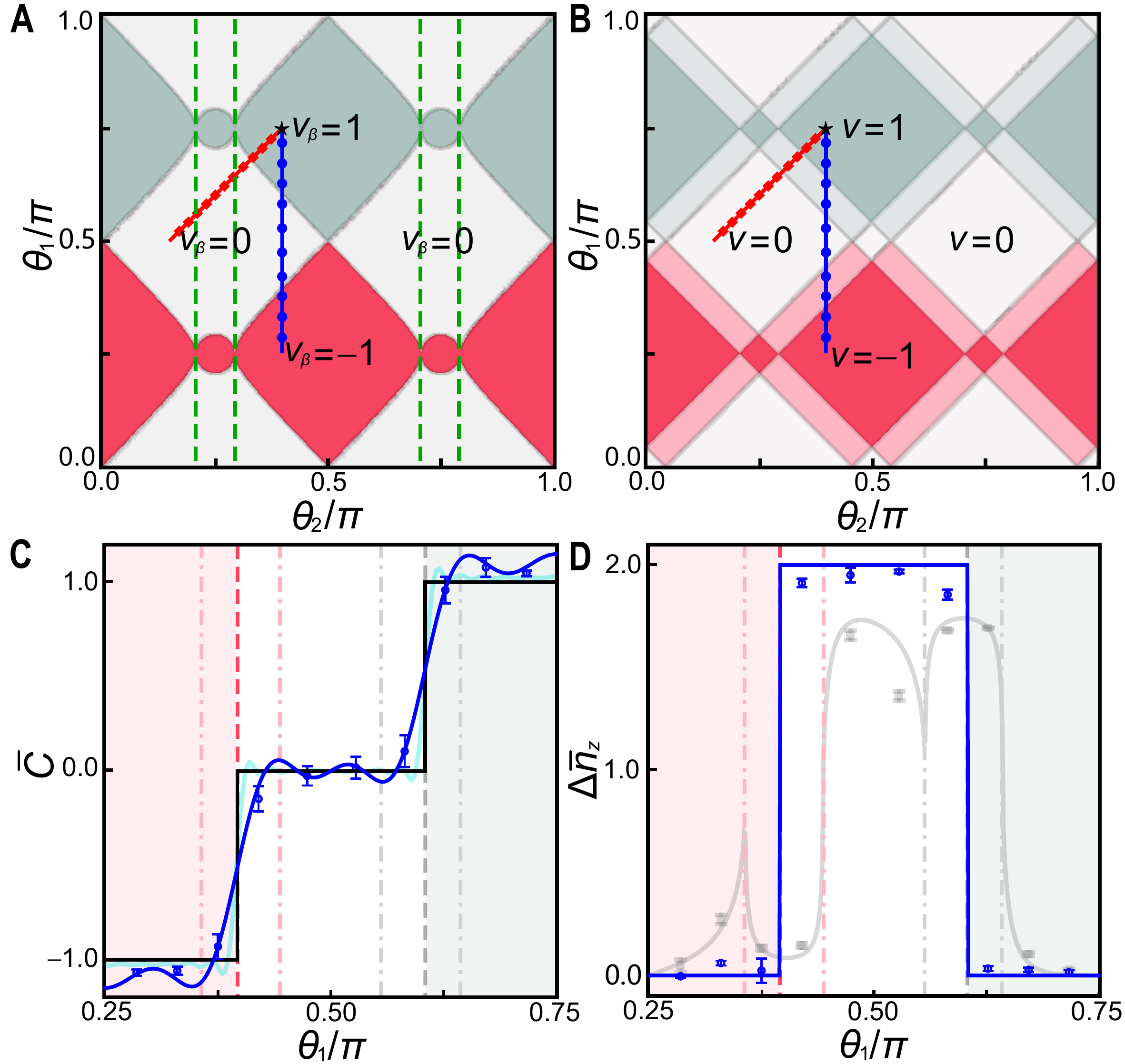}
   \caption{Detecting non-Bloch winding numbers. (a) Topological phase diagrams under the non-Bloch and (b) Bloch band theories. Different topological phases are characterized by the non-Bloch (Bloch) winding numbers $\nu_\beta$ ($\nu$) as functions of the coin parameters $\left(\theta_1,\ \theta_2\right)$ with a fixed $\gamma=0.2746$.
   Symbols along the blue and red lines indicate coin parameters we adopt along the two different paths on the phase diagram. While the GBZs along the blue line remain the same, those along the red line differ in radius.
   Black stars correspond to the coin parameters of $U^i$ for the quench processes. Green dashed lines in (a) indicate locations of the non-Bloch exceptional points.
   (c) The measured average chiral displacement $\bar{C}$ (blue symbols) for six-step quantum walks, along the blue path in the phase diagram with a fixed $\theta_2=0.4\pi$. The vertical dashed and dash-dotted lines indicate the locations of non-Bloch and Bloch topological phase boundaries, respectively. The blue (cyan) solid line corresponds to numerical simulations for six-step (twenty-step) quantum walks, and the solid black line indicates the numerically calculated $\nu_\beta$.
   (d) The measured $\Delta \bar{n}_z$ (blue symbols) along the blue path for six-step quantum walks. The blue solid line represents results from numerical simulations with the same number of steps. The gray symbols and lines show the experimental and numerical results of $\Delta \bar{n}_z$ respectively, calculated using the Bloch-band theory. Error bars are due to the statistical uncertainty in photon-number-counting.}
	\label{fig:windingE}
\end{figure}

{\it A non-Hermitian quantum walk.---}Our experiment is based on a single-photon quantum-walk configuration along a one-dimensional lattice, with the Floquet operator~\cite{photonskin,EP}
\begin{align}
U=R(\theta_{1})S_{2}R(\theta_{2})MR(\theta_{2})S_{1}R(\theta_{1}).
\label{U}
\end{align}
Here the shift operators $S_{1} =\sum_{x}|x\rangle\langle x|\otimes|\uparrow\rangle\langle\uparrow|+|x+1\rangle\langle x|\otimes|\downarrow\rangle\langle \downarrow|$ and $S_{2}  =\sum_{x}|x-1\rangle\langle x|\otimes|\uparrow\rangle\langle \uparrow|+| x\rangle\langle x|\otimes|\downarrow\rangle\langle\downarrow|$, and the rotation operator $R(\theta) =\one_\text{w} \otimes e^{-i \theta \sigma_{y}}$ with $\one_\text{w}=\sum_x|x\rangle\langle x|$. The lattice sites are labelled by $x$, and the internal states $\left|\uparrow\right\rangle$ and $\left|\downarrow\right\rangle$ are eigenstates of the Pauli matrix $\sigma_z$, which are encoded in the horizontal and vertical polarizations of photons, respectively. The gain-loss operator, $M=\one_\text{w} \otimes e^{\gamma\sigma_z}$, is related to the experimentally implemented selective-loss operator
$M_E=\one_\text{w}\otimes \left(\left|\uparrow\right\rangle\left\langle \uparrow\right|+\sqrt{1-p}\left|\downarrow\right\rangle\left\langle \downarrow\right|\right)$ through the map $M=e^\gamma M_E$ with $\gamma=-\frac{1}{4}\ln (1-p)$ ($0\leq p\leq 1$). All elements of the Floquet operator $U$ are implemented using a combination of wave plates, beam displacers and partially polarizing beam splitters~\cite{photonskin,EP,suppl}.

\begin{figure*}
\includegraphics[width=0.8\textwidth]{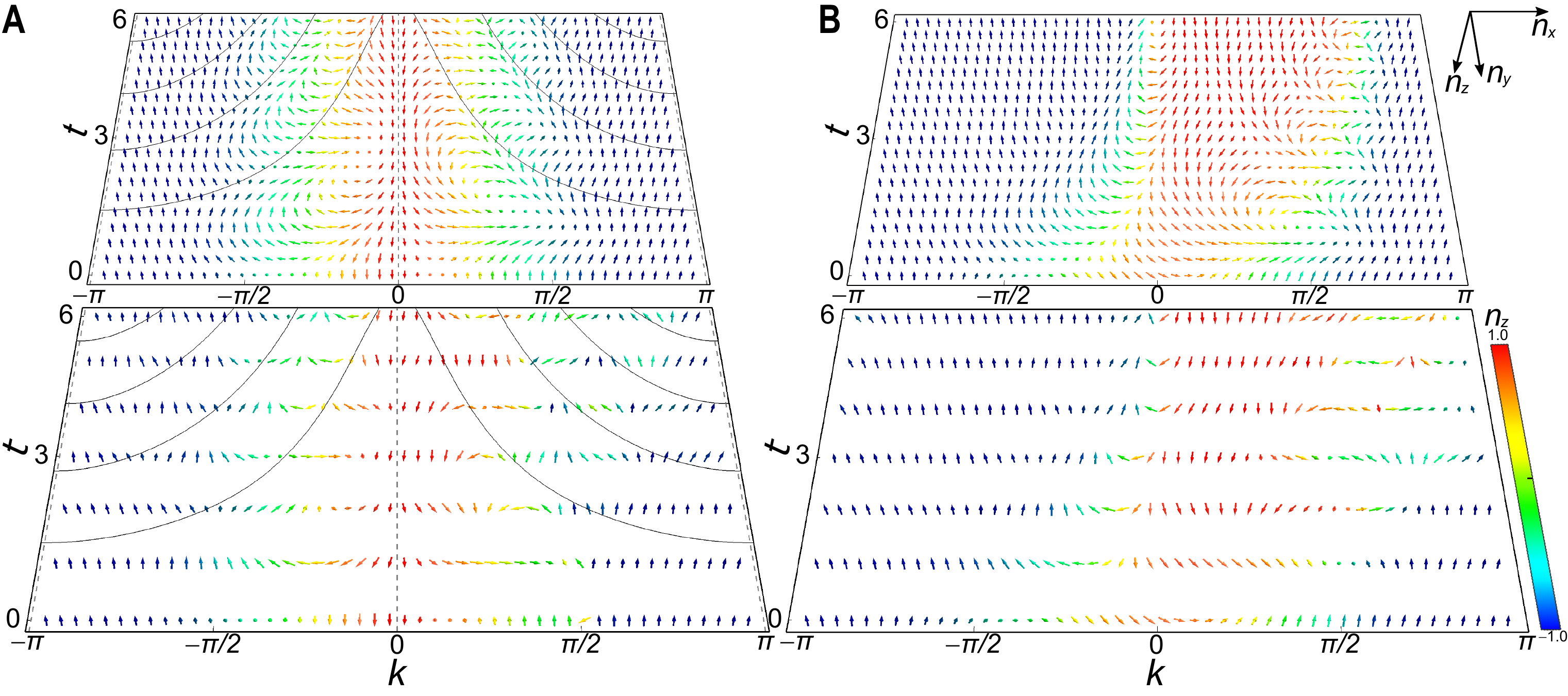}
   \caption{Experimental (lower layer) and numerical results (upper layer) of dynamic spin textures. (a) The dynamic spin textures $\bm{n}(k_{\beta},t)$  for the same quench process projected into the GBZ and (b) BZ.
   For either case, the quench takes place between $\nu^i=1$ and $\nu^f=0$, with $U^i$ characterized by $\left(\theta_1=3\pi/4,\ \theta_2=0.4\pi\right)$, and $U^f$ by $\left(\theta_1=0.528\pi,\ \theta_2=0.4\pi\right)$. Spin textures are colored according to the value of $n_z(k_\beta,t)$. For the non-Bloch dynamics, the momentum-dependent periods of the oscillatory spin dynamics are illustrated by the solid black lines in (a). By contrast, spin dynamics in (b) are non-oscillatory, due to the complex quasienergy spectra.
   }
	\label{fig:spinT}
\end{figure*}

{\it Non-Bloch topological invariants in non-Hermitian quantum walks.---}The Floquet operator $U$ is topological, and features the non-Hermitian skin effect in the presence of boundaries. Following the non-Bloch band theory~\cite{WZ1,WZ2,murakami}, its topology under an open boundary condition is characterized by non-Bloch winding numbers $\nu_\beta$ defined on the GBZ~\cite{suppl}.
Specifically, this is achieved by replacing the Bloch phase factor
$e^{ik}$ with a factor $\beta=r e^{ik_\beta}$ representing the GBZ, which manifests itself as a closed loop on the complex plane. Note that the GBZ of $U$ under the open boundary condition is always circular with a radius $r$, and is parameterized by $k_\beta \in \left[0,2\pi\right)$, regarded as the generalized quasimomentum in the non-Bloch band theory.
The spirit of the substitution is to take into account the deviation of the nominal bulk eigenstates (encoded in $\beta$) from the extended Bloch waves (encoded in $e^{ik}$). Crucially, only the non-Bloch winding number, rather than its Bloch counterpart, is capable of restoring the bulk-boundary correspondence under the open boundary condition~\cite{WZ1,photonskin}.

Further, as $U$ respects the chiral symmetry with $\Gamma U\Gamma=U^{-1}$ and $\Gamma=\one_\text{w}\otimes\sigma_x$, the non-Bloch winding number is reflected in the time-averaged biorthogonal chiral displacement~\cite{ZCW,suppl}
\begin{align}
\bar{C}=\frac{2}{T}\sum^T_{t=1}\text{Re} \Big(\langle\chi(t)|\Gamma X|\psi(t)\rangle\Big),
\label{eq:Ct}
\end{align}
provided that the eigenenergy spectra of the effective Hamiltonian are completely real. Here the effective Hamiltonian $H$ is defined through $U=e^{-iH}$, $X$ is the spatial position operator, $T$ is the total number of discrete time steps, $|\psi(t)\rangle=U^t|\psi(0)\rangle$ and $|\chi(t)\rangle=\left[(U^{-1})^\dag\right]^t|\psi(0)\rangle$, with $|\psi(0)\rangle=|0\rangle\otimes |\uparrow\rangle$.
Given the presence of the non-Bloch PT symmetry, $\bar{C}$ offers an experimentally feasible access to the non-Bloch winding number.

Alternatively, the non-Bloch winding number can be inferred from the dynamic spin structures following a quantum quench. Such a quench-based detection is hinged on the understanding that a quantum walk constitutes a stroboscopic simulation of the time evolution driven by the effective Hamiltonian $H$. Initializing the walker in an eigenstate of an initial Floquet operator $U^i$, and evolving it under a final Floquet operator $U^f$, we simulate the quench dynamics between the initial and final effective Hamiltonians $H^i$ and $H^f$.

In a previous experiment with non-unitary photonic quantum walks, Bloch winding numbers were measured through dynamic quench processes driven by non-Hermitian effective Hamiltonians devoid of skin effect~\cite{XPNC}. It was shown that a dynamic skyrmion structure $\bm{n}(k,t)$ emerges in the quasimomentum-time domain, as long as the initial and final effective Hamiltonians possess different winding numbers, while both in the exact PT phase with completely real quasienergy spectra. Thus, once the topology of $H^i$ is known, for instance through state preparation, one can readily deduce the topology of $H^f$ from the measured dynamic spin texture~\cite{Chen17,Ueda17,iS,XPNC}.

To detect non-Bloch topological invariants, however, the quench-based scheme above cannot be directly applied, not least because the Bloch wave vectors $k$ are no longer good quantum numbers under open boundaries and non-Hermitian skin effects. Instead, we project the quench dynamics onto the GBZ, to visualize the dynamic spin texture $\bm{n}(k_\beta,t)$ in the generalized quasimomentum-time domain, now parameterized by ($k_\beta$,$t$)~\cite{prr}. Here $\bm{n}(k_\beta,t)$ fully captures the dynamics of the projected density matrix on the GBZ, and is a real unit vector that can be visualized on the Bloch sphere ~\cite{suppl}.
Within the framework of this non-Bloch quench description, $\bm{n}(k_\beta,t)$ acquires a skyrmion structure when $H^i$ and $H^f$ have distinct non-Bloch winding numbers while both possess completely real quasienergy spectra, now protected by the non-Bloch PT symmetry.

{\it Detecting non-Bloch topological invariant.---}We first measure the time-averaged chiral displacement $\bar{C}$. Initializing the walker in the state $\left|0\right\rangle\otimes \left|\uparrow\right\rangle$, we evolve it with $U$ and $(U^{-1})^\dag$, respectively, and perform state tomography to reconstruct the auxiliary density matrices $\rho_R=|\psi(t)\rangle\langle \psi(t)|$ and $\rho_L=|\chi(t)\rangle\langle \chi(t)|$. The time-averaged chiral displacement is then constructed through
\begin{align}
\bar{C}=\frac{2}{T}\sum^T_{t=1}\text{Re}\Big[\text{Tr} (\Gamma X  \rho_R\rho_L)\Big].
\end{align}
In Fig.~\ref{fig:windingE}(c), we show the measured time-averaged chiral displacements for quantum walks with different coin parameters along the blue path in Fig.~\ref{fig:windingE}(a). The measured $\bar{C}$ agrees well with the non-Bloch phase diagram, oscillating around the quantized value of the non-Bloch winding number within each phase. At the phase boundaries, jumps in the measured $\bar{C}$ are clearly identified.

We complement the measurements above using the non-Bloch quench dynamics. Choosing an initial $U^i$ characterized by $(\theta_1=3\pi/4,\ \theta_2=0.4\pi)$ (black star in the phase diagram), we prepare the initial state $\left|\psi^i\right\rangle=0.471\left|0\right\rangle\otimes\left(\left|\uparrow\right\rangle+\left|\downarrow\right\rangle\right)+0.667\left|-1\right \rangle\otimes\left|\uparrow\right\rangle+0.236\left|-2\right\rangle\otimes\left(-\left|\uparrow\right\rangle+\left|\downarrow\right\rangle\right) $, by subjecting the walker localized at $x=0$ to preliminary gate operations.
Note that the initial state is an eigenstate of $U^i$ regardless of boundary conditions. This is expected as such a quasi-local initial state is entirely in the bulk and should be independent of boundary conditions.
The initial state is then quenched under final Floquet operators chosen along the blue path in Fig.~\ref{fig:windingE}(a).
Following Ref.~\cite{prr}, we define
\begin{align}
\Delta \bar{n}_z=\frac{1}{T}\sum_{t=0}^T \left[n_z(0,t)-n_z(\pi,t)\right].
\end{align}
By capturing the spin-texture difference at $k_\beta=0$ and $k_\beta=\pi$ in the GBZ,
$\Delta\bar{n}_z$ typically reflects the emergence/disappearance of dynamic skyrmions.

As illustrated in Fig.~\ref{fig:windingE}(d), abrupt changes in the measured $\Delta\bar{n}_z$ indicate locations of the non-Bloch phase transitions, which are consistent with the measurements of chiral displacement.
By contrast, $\Delta\bar{n}_z$ calculated using the Bloch vectors ($k$ in the BZ) is typically not quantized.
Here two remarks are in order.

First, when projected onto the BZ, $\Delta\bar{n}_z$ also exhibits discontinuities at the Bloch phase boundaries [the gray solid line in Fig.~\ref{fig:windingE}(d)]. However, dynamic skyrmions do not exist in the quasimomentum-time domain under the Bloch-band description. This is more clearly seen by explicitly mapping out the spin texture $\bm{n}(k_\beta,t)$, as shown in Fig.~\ref{fig:spinT}(b). Notably, $\bm{n}(k,t)$ under the Bloch-band theory is steady-state approaching, with $\displaystyle{\lim_{t\rightarrow +\infty}}\bm{n}(k,t)=(0,0,\pm1)$ for each $k$ sector. This derives from the spectral topology of the non-Hermitian skin effect~\cite{kawabataskin,fangchenskin}, which necessitates a complex quasienergy spectra of both $H^{i}$ and $H^f$ under the periodic boundary condition.

Second, in Fig.~\ref{fig:windingE}(d), as $U^i$ is characterized by $\theta^i_1=3/4\pi$ with $\nu^i_\beta=1$, the measured $\Delta \bar{n}_z$ is $2$ when the non-Bloch winding number $\nu_\beta^f$ of $U^f$ differs from $\nu^i_\beta$ by an odd integer. When $\nu^i_\beta$ and $\nu^f_\beta$ differ by a nonzero even integer, the measured $\Delta \bar{n}_z$ is always zero. This is due to the proliferation of dynamic skyrmions in the generalized quasimomentum-time domain, such that $k_\beta=0$ and $k_\beta=\pi$ no longer underpin the same skyrmion structure. Importantly, it is
qualitatively different from the case with $\nu_\beta^i=\nu_\beta^f$ where no skyrmions exist. One can always differentiate these two cases by explicitly mapping out the dynamic spin textures~\cite{suppl}. Since adjacent phases in Fig.~\ref{fig:windingE}(a) differ in the non-Bloch winding number only by $1$, when sweeping parameters of $U^f$ along a path, a jump between quantized values of $\Delta \bar{n}_z$ is always indicative of a non-Bloch topological phase transition.

\begin{figure}[t!]
\includegraphics[width=0.45\textwidth]{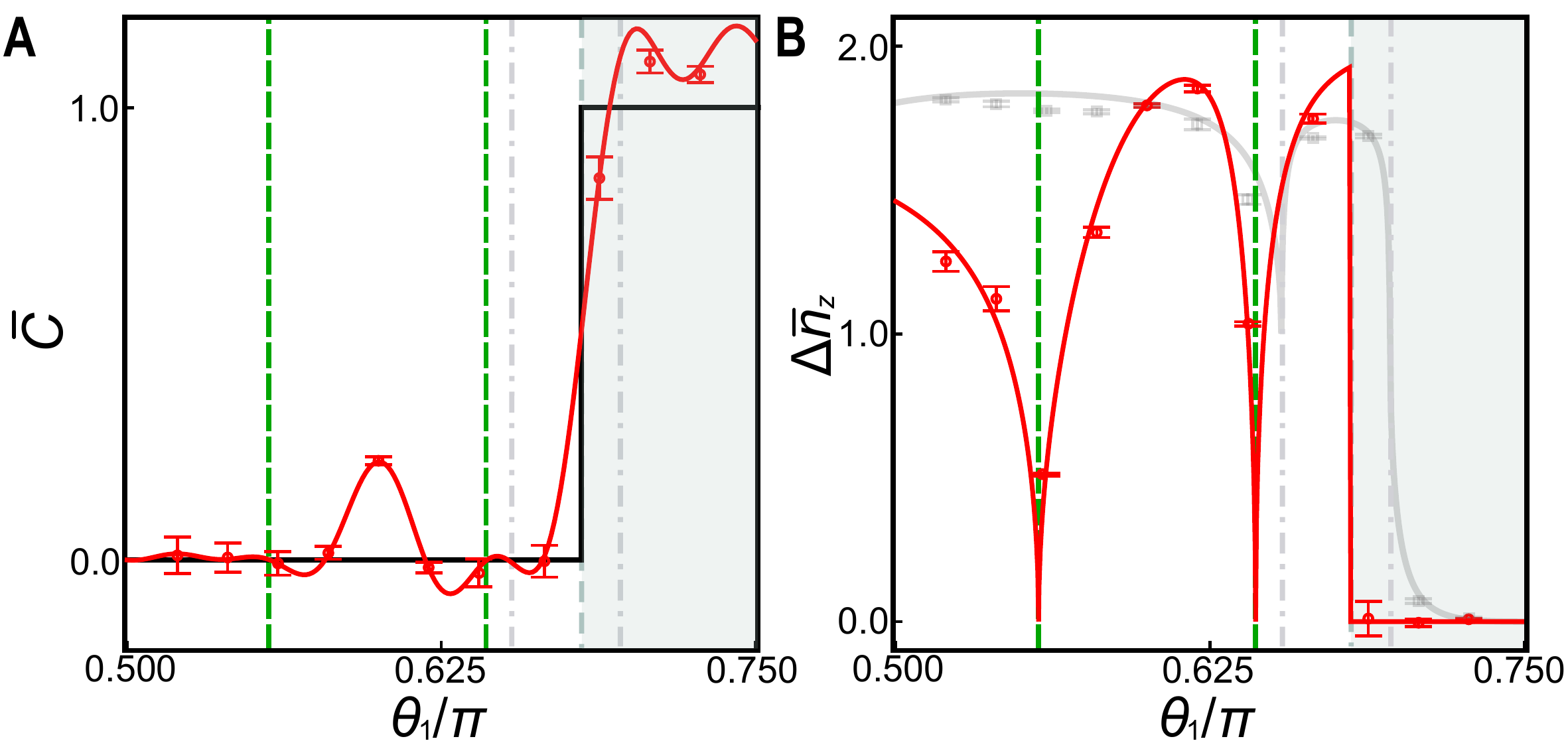}
   \caption{ Measurements across non-Bloch exceptional points. (a) Measured $\bar{C}$ and (b) $\Delta \bar{n}_z$ for six-step quantum walks, with varying $\theta_1$ along the red lines in Fig.~\ref{fig:windingE}. Theoretical results are shown in red solid lines, and experimental results are presented by red symbols. The vertical dashed and dash-dotted gray lines indicate, respectively, the locations of non-Bloch and Bloch topological phase boundaries. The vertical dashed green lines indicate the locations of the non-Bloch exceptional points. The solid black line in (a) denotes the numerically calculated non-Bloch winding number. The gray solid line and squares in (b) indicate the theoretical and measured $\Delta \bar{n}_z$ according to the Bloch band theory. Error bars are due to the statistical uncertainty in photon-number counting.}
	\label{fig:windingUE}
\end{figure}

\begin{figure}[htp]
\includegraphics[width=0.45\textwidth]{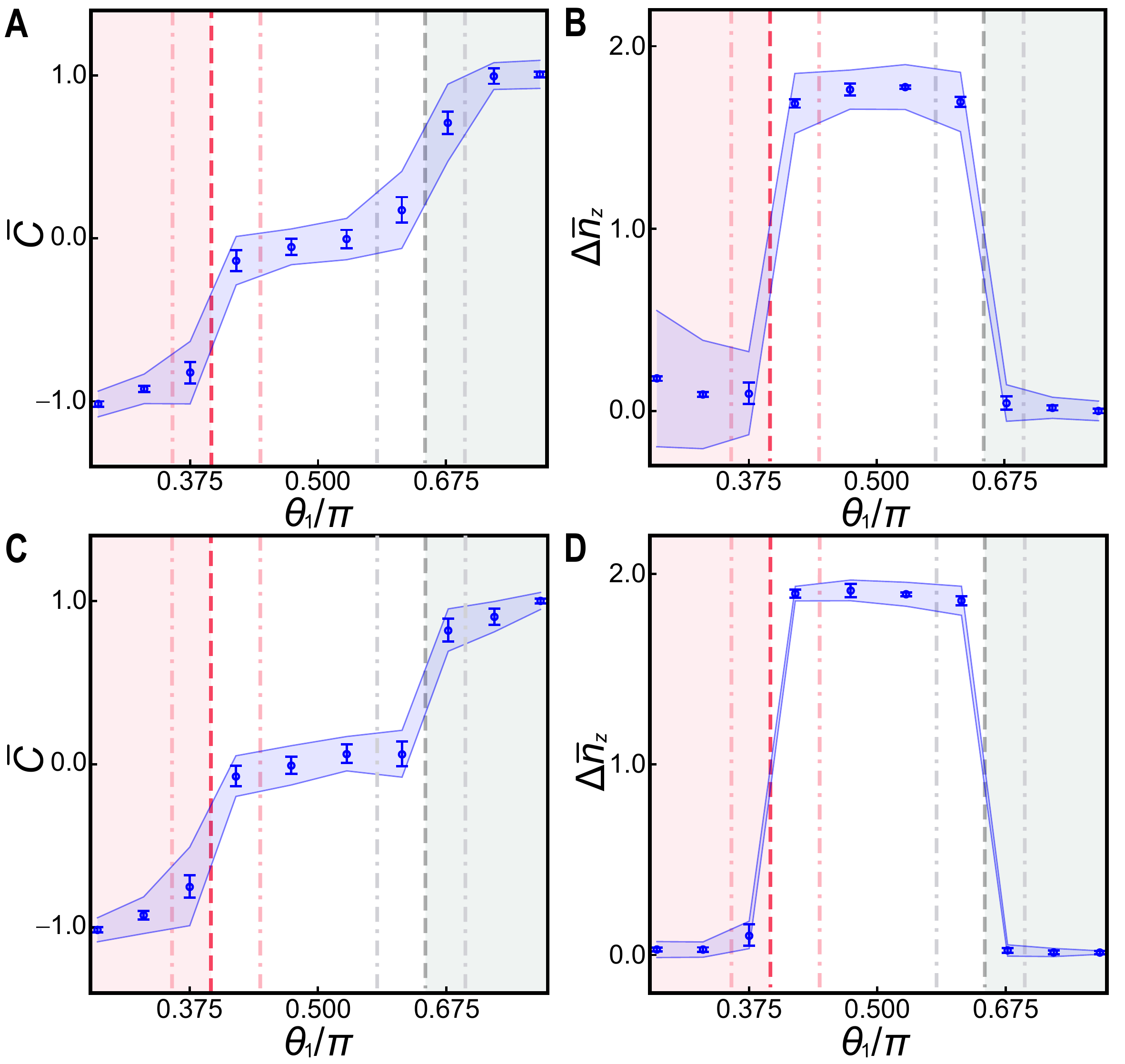}
   \caption{Impact of disorder. (a) Time- and disorder-averaged biorthogonal chiral displacement and (b) $\Delta \bar{n}_z$ for six-step quantum walks under static disorder. The coin parameters are randomly chosen in the interval $\left[\theta_i-\pi/20,\theta_i+\pi/20\right]$ ($i=1,2$) at each spatial location, where $\theta_1$ is scanned along the blue line in Fig.~\ref{fig:windingE}(a). Coin parameters do not change with time steps. (c) Time- and disorder-averaged biorthogonal chiral displacement and (d) $\Delta \bar{n}_z$ for six-step quantum walks under dynamic disorder. The symbols and shaded regions respectively indicate the mean values of experimental measurements and the range of the standard deviations averaged over $10$ different disorder configurations for each set of $(\theta_1,\theta_2)$. The vertical lines indicating the non-Bloch and Bloch phase boundaries are the same as those in Figs.~\ref{fig:windingE}(c) and (d).
   }
	\label{fig:disorder}
\end{figure}

{\it Robustness test of non-Bloch topological invariants.---}In the experiments above, we measure the non-Bloch topological invariants along a path where the GBZ remains the same, with completely real quasienergy spectra under the open boundary condition.
To demonstrate the generality of our schemes, we choose an alternative path [red line in Fig.~\ref{fig:windingE}(a)], along which the GBZ, though still circular, has a varying radius. Further, the system under the open boundary condition features non-Bloch PT transitions along the path, with the eigenvalues of $H$ being real only in the exact PT region. As shown in Fig.~\ref{fig:windingUE}, both the measured $\bar{C}$ and $\Delta \bar{n}_z$ correctly reflect the non-Bloch phase boundary. Remarkably, $\bar{C}$ and $\Delta \bar{n}_z$ appreciably deviate from quantized values in the non-Bloch PT broken region, with $\Delta \bar{n}_z$ exhibiting non-analytic behavior at the exceptional points separating the exact and PT broken phases(vertical green dashed lines)~\cite{YangLanPT}. The post-quench dynamic spin texture thus offers a useful probe for the non-Bloch exceptional points and the associated criticality.

Finally, we introduce symmetry-preserving disorder to $U^f$, and investigate the robustness of the non-Bloch topological invariants, as well as that of our detection schemes. Two different types of disorder are studied~\cite{ZXB17}. First, static disorder is introduced by randomly modulating both coin parameters around their mean values in the range of $\left[-\pi/20,\pi/20\right]$. The disorder depends on spatial location but does not change from step to step. We also impose dynamic disorder by randomly modulating the coin parameters within the same interval as above, but now the disorder changes from step to step while being spatially homogeneous. As demonstrated in Fig.~\ref{fig:disorder}, the disorder-averaged $\bar{C}$ and $\Delta \bar{n}_z$ still show marked difference in different non-Bloch topological phases, confirming both the robustness of the non-Bloch topological invariants and that of our detection schemes against disorder.

{\it Discussion.---}We report the direct detection of non-Bloch topological invariants from dynamic signals in the bulk, thus establishing them as observable quantities in the thermodynamic limit. Consistent with the theoretical characterization of non-Bloch topological invariants,
our experiment is hinged upon the concept of non-Bloch band theory, where the application of the GBZ is essential.
The experiment thus serves to further underline the importance of the non-Bloch band theory in providing a coherent understanding regarding lattice models with non-Hermitian skin effects.
As both of our detection schemes are sensitive to the non-Bloch PT symmetry, they also provide dynamic probes to the non-Bloch exceptional points.
For future studies, it would be desirable to devise detection schemes for non-Bloch topological invariants in models of higher dimensions. Our work paves the way for direct experimental study of non-Bloch topology.

\section*{Acknowledgments}
We thank Zhong Wang for reading the manuscript and helpful comments.
This work has been supported by the National Natural Science Foundation of China (Grant Nos. 12025401, U1930402 and 11974331). W. Y. acknowledges support from the National Key Research and Development Program of China (Grant Nos. 2016YFA0301700 and 2017YFA0304100). L. X. acknowledges support from the Project Funded by China Postdoctoral Science Foundation (Grant Nos. 2020M680006 and 2021T140045).


\clearpage
\appendix

\widetext

\begin{center}
\textbf{\large  Supplemental Material for  ``Detecting Non-Bloch Topological Invariants in Quantum Dynamics''}
\end{center}

\section{Reconstruction of time-evolved states}
To simplify the reconstruction of the time-evolved states in our experiment, we assume that the system is always in a pure state. The wave function at a certain time step $t$ is therefore
\begin{equation}
\ket{\Psi(t)}=\sum_{x=-\mathcal{N}}^{\mathcal{N}}\ket{x}\otimes\ket{\psi_x(t)},
\label{eq:pureS}
\end{equation}
where $x\in\left[-\mathcal{N},\ \mathcal{N}\right]$ labels the positions of the walker on a spatial lattice with a total of $2\mathcal{N}+1$ sites. For each position $x$, the internal coin state can be written as
\begin{equation}
\ket{\psi_x(t)}=e^{i\varphi_{x,t}}\left(a_{x,t}\ket{\uparrow}+e^{i\delta_{x,t}}b_{x,t}\ket{\downarrow}\right)
\label{eq:coinS}
\end{equation}
with $\varphi_{x,t},\ \delta_{x,t}\in\left[0,2\pi\right)$ and $a_{x,t},\ b_{x,t}\in\mathbb{R}$. The internal states $|\uparrow\rangle$ and $|\downarrow\rangle$ represent the horizontal and vertical polarizations of photons, respectively.

\begin{figure}[htp]
\includegraphics[width=0.5\textwidth]{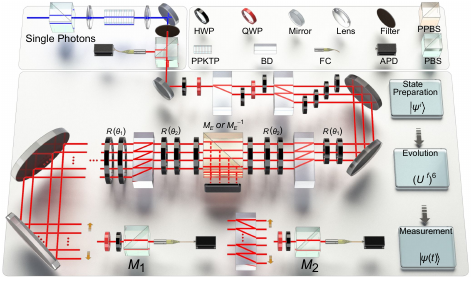}
\caption{Experiment setup. Heralded single photons generated via the spontaneous parametric down conversion in periodically poled potassium titanyl phosphate crystal (PPKTP) are encoded as the walker. The initial quasi-localized state is prepared by passing the single photons through a polarizing beam splitter (PBS), two beam displacers (BDs), two quarter-wave plates (QWPs) at $0$ and seven half-wave plates (HWPs). The selective-loss operator, coin rotation and shift operators in the Floquet operator $U$ are realized by a partially polarizing beam splitter (PPBS) combined with two HWPs, two HWPs with certain setting angles and a BD, respectively. To reconstruct the evolved states $\ket{\Psi(t)}$, two kinds of measurements, local projective measurements with and without an extra shift operator $S_2$, are applied before the photons are detected by avalanche photodiodes (APDs).}
\label{fig:setup}
\end{figure}

As illustrated in Fig.~\ref{fig:setup}, our experiments involve two kinds of measurements.
First, using a quarter-wave plate (QWP), a half-wave plate (HWP) and a polarizing beam splitter (PBS), we perform projective measurements on each position in the basis
\begin{align}
P^H=&\ketbrad{\uparrow} , \\
P^V=&\ketbrad{\downarrow} , \\
P^R=&\frac{1}{2}\left(\ket{\uparrow}-i\ket{\downarrow}\right)\left(\bra{\uparrow}+i\bra{\downarrow}\right) , \\
P^D=&\frac{1}{2}\left(\ket{\uparrow}+\ket{\downarrow}\right)\left(\bra{\uparrow}+\bra{\downarrow}\right).
\label{eq:proj1}
\end{align}
The photon coincidences measured in the four basis are denoted as $\{N^H_x(t),\ N^V_x(t),\ N^R_x(t),$\ $ N^D_x(t)\}$, which satisfy the following relations
\begin{align}
N^H_x(t)=&\mathcal{M}a_{x,t}^2, \\
N^V_x(t)=&\mathcal{M}b_{x,t}^2, \\
N^R_x(t)=&\mathcal{M}\frac{a_{x,t}^2+b_{x,t}^2-2a_{x,t}b_{x,t}\text{sin}\delta_{x,t}}{2}, \\
N^D_x(t)=&\mathcal{M}\frac{a_{x,t}^2+b_{x,t}^2+2a_{x,t}b_{x,t}\text{cos}\delta_{x,t}}{2},
\end{align}
where $\mathcal{M}$ represents the total coincidence counts input to the initial state preparation. Thus, we obtain a set of $4(2\mathcal{N}+1)$ counts after these measurements.

After inserting an extra beam displacer (BD) to achieve the shift operator $S_2$, the horizontally polarized photons in the spatial modes $x+1$ and the vertically polarized photons in the spatial modes $x$ would be injected into the same spatial modes $x$. The coin state in positions $x\in\left[-\mathcal{N},\ \mathcal{N}-1\right]$ are therefore changed into
\begin{equation}
\ket{\psi^\prime_x(t)}=e^{i\varphi_{x+1,t}}a_{x+1,t}\ket{\uparrow}+e^{i\left(\varphi_{x,t}+\delta_{x,t}\right)}b_{x,t}\ket{\downarrow}.
\end{equation}
We then apply the projective measurement $\{P^R,\ P^D\}$ at different positions with a QWP, a HWP and a PBS. We denote the obtained photon coincidences as $\{\tilde{N}^R_x(t),\ \tilde{N}^D_x(t)\}$, which satisfy the relations
\begin{small}
\begin{align}
\tilde{N}^R_x(t)=&\mathcal{M}\frac{a_{x+1,t}^2+b_{x,t}^2+2a_{x+1,t}b_{x,t}\text{sin}\left(\varphi_{x+1,t}-\varphi_{x,t}-\delta_{x,t}\right)}{2}, \\
\tilde{N}^D_x(t)=&\mathcal{M}\frac{a_{x+1,t}^2+b_{x,t}^2-2a_{x+1,t}b_{x,t}\text{cos}\left(\varphi_{x+1,t}-\varphi_{x,t}-\delta_{x,t}\right)}{2}.
\label{eq:proj2}
\end{align}
\end{small}
We then get a set of $2(2\mathcal{N})$ counts during this step.

Since the phase $\varphi_{\mathcal{N},t}$ is unimportant, the total number of photon counts we have $4(3\mathcal{N}+1)$ is sufficiently large to obtain all the parameters $\{\varphi_{x,t},\ \delta_{x,t},\ a_{x,t},\ b_{x,t}\}$. Following the conventional methods, we optimize these parameters by finding the minimum of the ``likelihood'' function
\begin{equation}
\mathcal{L}=\sum_{x=-\mathcal{N}}^{\mathcal{N}}\sum_{i\in\{H,V,R,D\}}\frac{\left[N^{i}_x(t)^\prime-N^i_x(t)\right]^2}{2N^i_x(t)}+\sum_{x=-\mathcal{N}}^{\mathcal{N}-1}\sum_{i\in\{R,D\}}\frac{\left[\tilde{N}^i_x(t)^\prime-\tilde{N}^i_x(t)\right]^2}{2\tilde{N}^i_x(t)},
\end{equation}
where $N^{i}_x(t)^\prime$ and $\tilde{N}^i_x(t)^\prime$ are the measured photon counts.

\begin{figure}[htp]
\centering
\includegraphics[width=0.45\textwidth]{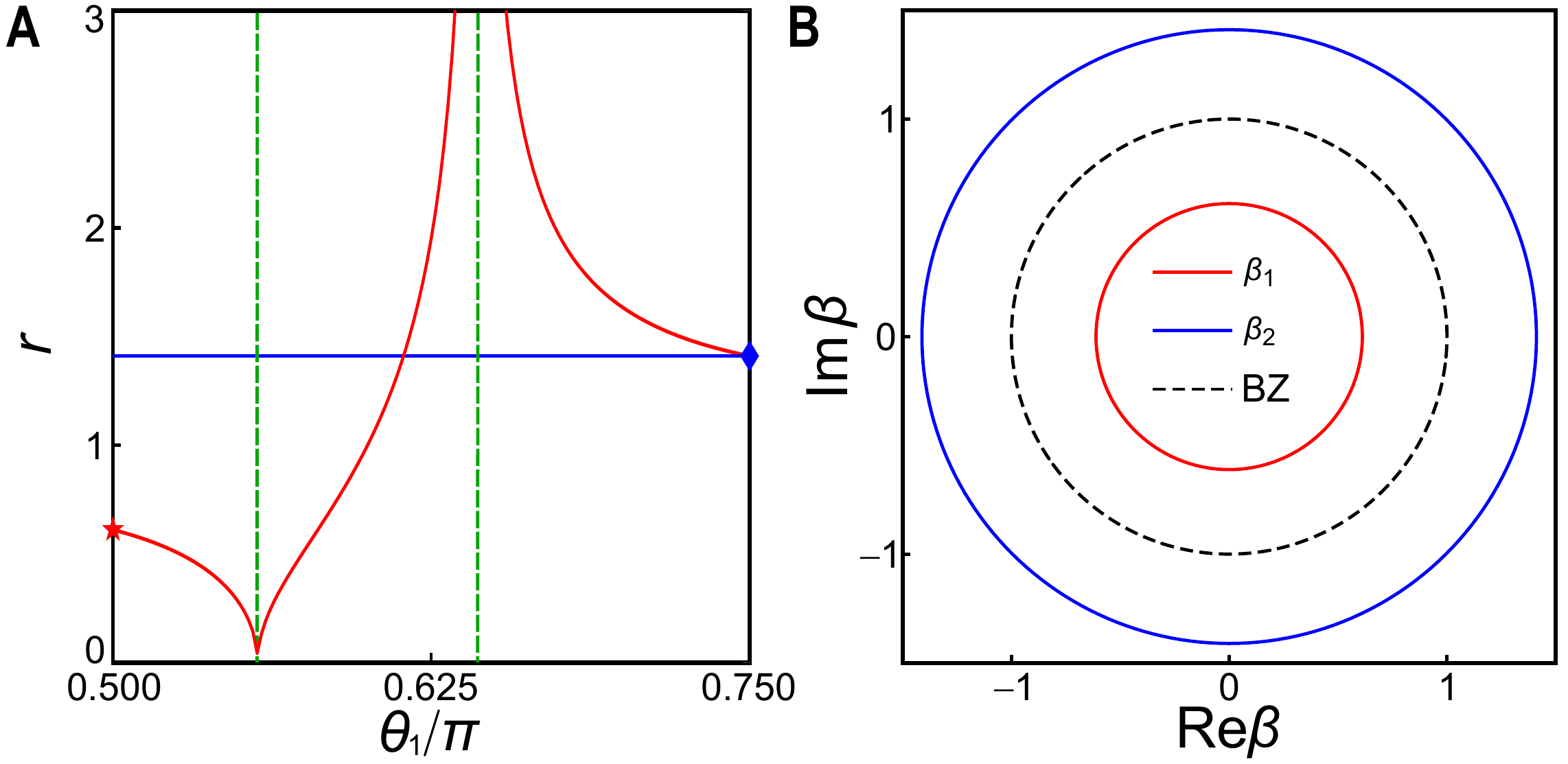}
\caption{(a) Variation of the GBZ radius along the blue and red paths (lines with matching color) in Fig.~\ref{U}(a) of the main text. Vertical green dashed lines indicate locations of the non-Bloch exceptional point.
(b) GBZs on the complex plane. Red and blue circles respectively correspond to the GBZs with $r=0.61$ [star in (a)] and $r=1.41$ [diamond in (a)]. The black circle corresponds to the BZ with a radius $r=1$.
}
\label{fig:Sgbz}
\end{figure}

\begin{figure*}[htp]
\centering
\includegraphics[width=\textwidth]{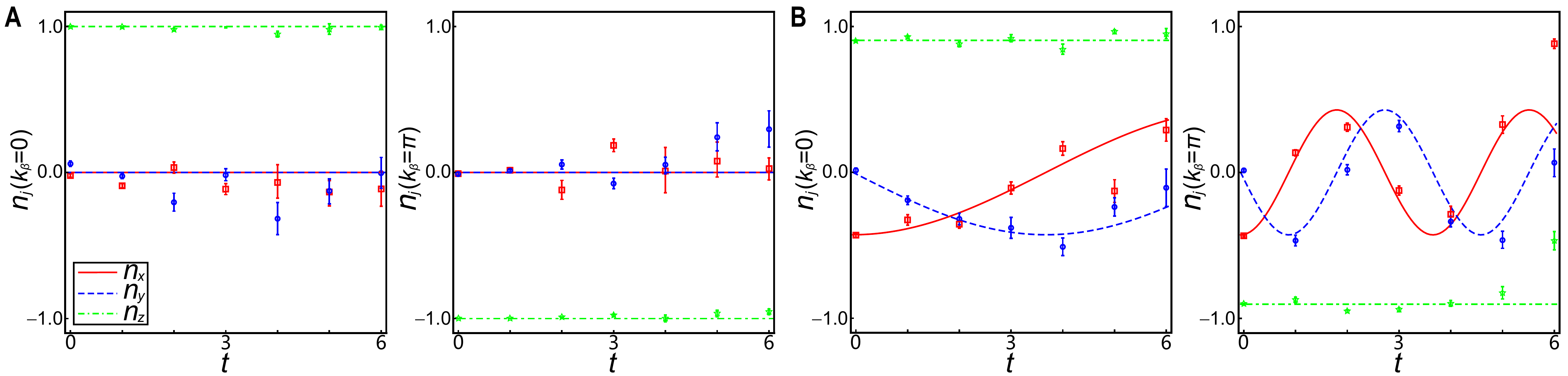}
\caption{Time-evolution of $\bm{n}(k_\beta,t)$ for $k_\beta\in \left[0,\ \pi\right]$ in the GBZ. The initial Floquet operator is characterized by $\left(\theta_1=3\pi/4,\ \theta_2=0.4\pi\right)$ with $\nu^i_\beta=1$ and $r_i=1.41$.
(a) $U^f$ is characterized by $\left(\theta_1=0.528\pi,\ \theta_2=0.4\pi\right)$ with
$\nu^f_\beta=0$ and $r_f=1.41$ (same parameters as in Fig.~\ref{eq:Ct}(a) of the main text). Fixed points rigorously exist at $k_\beta=0$ and $k_\beta=\pi$ in this case.
(b) $U^f$ is characterized by $\left(\theta_1=0.666\pi,\ \theta_2=0.316\pi\right)$ with
$\nu^f_\beta=0$ and $r_f=2.23$ (same as Fig.~\ref{fig:S3}(a)). In this case, fixed points do not exist.
Numerical results are presented by solid curves, and experimental data are shown in symbols. Error bars are due to the statistical uncertainty in photon-number counting.}
\label{fig:fp}
\end{figure*}

\begin{figure*}[htp]
\centering
\includegraphics[width=0.8\textwidth]{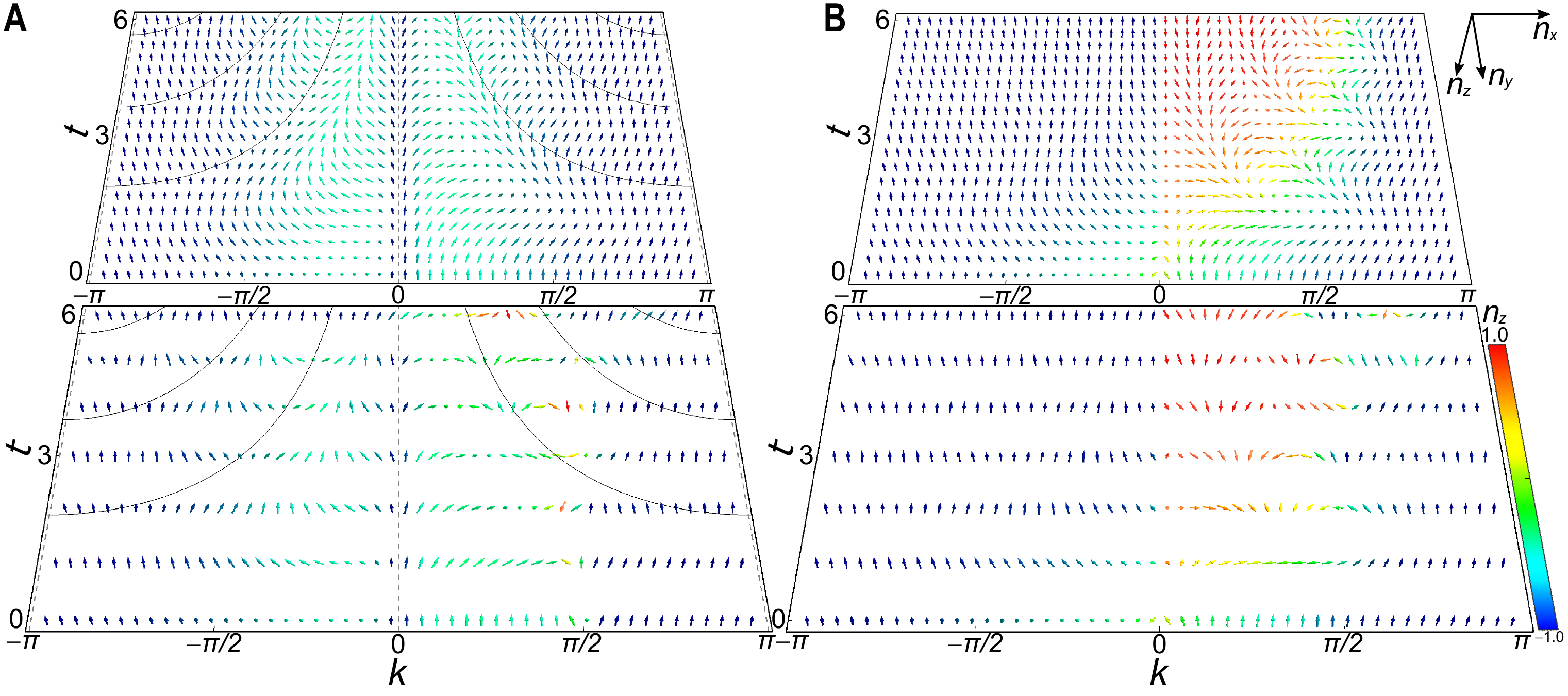}
\caption{(a) Dynamic spin textures $\bm{n}(k_{\beta},t)$ for a quench process between non-Bloch topological phases with the same non-Bloch winding number and the same GBZ radius $r_i=r_f=1.41$. The initial Floquet operator $U^i$ is characterized by $\left(\theta_1=3\pi/4,\ \theta_2=0.4\pi\right)$ with $\nu^i_\beta=1$, and $U^f$ is characterized by $\left(\theta_1=0.627\pi,\ \theta_2=0.4\pi\right)$ with $\nu^f_\beta=1$. (b) Dynamic spin texture $\bm{n}(k,t)$ of the same quench process as in (a), but projected into the BZ using the Bloch band theory.
Here $\nu^i=1$ and $\nu^f=1/2$.
Experimental and numerical results are shown in the upper and lower layers, respectively.
The spin textures are colored according to the value of $n_z(k,t)$. For the non-Bloch dynamics, the momentum-dependent periods of the oscillatory spin dynamics are illustrated by the solid black lines in (a). Both $U^i$ and $U^f$ are in the non-Bloch exact PT phase, and fixed points of the dynamics are shown by the dashed lines.
By contrast, spin dynamics in (b) are non-oscillatory, due to the complex quasienergy spectra.
}
\label{fig:S1}
\end{figure*}

\section{Non-Bloch winding number}
The quantum walk driven by $U$ is topological, with its winding number given by
\begin{align}
\nu=\frac{1}{2\pi}\int dk \frac{-d_z\partial_{k} d_y+d_y\partial_{k} d_z}{d_y^2+d_z^2},
\label{eq:nu}
\end{align}
where the Bloch vector $\bm{d}=\text{Tr} \left(H\bm{\sigma}\right)/|\text{Tr} \left(H\bm{\sigma}\right)|$ [$k\in\left[0,2\pi\right)$]. Here the effective Hamiltonian $H$ is defined through $U=e^{-iH}$, and $\bm{\sigma}=(\sigma_x,\sigma_y,\sigma_z)$ ($\sigma_{x,y,z}$ are the Pauli matrices).

More importantly, $U$ features the non-Hermitian skin effect in the presence of boundaries, toward which all eigenstates are exponentially localized~\cite{photonskin}. It follows that the topological properties of $U$ must be characterized using the non-Bloch band theory, with its topological invariants defined on the GBZ~\cite{WZ1}.
As discussed in the main text, the non-Bloch winding number $\nu_\beta$ is calculated by making the replacement $e^{ik}\rightarrow r e^{ik_\beta}$
in Eq.~(\ref{eq:nu}), and integrate over $k_\beta$.

We note that, as is the case with quantum walks (and Floquet topological systems in general), a pair of winding numbers can be defined for $U$ when it is cast in different time frames~\cite{asboth,XPNC}. Here, throughout our work, we only consider the winding number defined in the time frame fixed by Eq.~(\ref{U}) in the main text. Both of our detection schemes will also work for the other winding number if we perform our analysis in the alternative time frame.

\section{Calculating the generalized Brillouin zone}

In this section, we briefly outline the calculation of the generalized Brillouin zone (GBZ). Following Ref.~\cite{EP}, we rewrite the Floquet operator as
\begin{align}
U=\sum_{x}|x\rangle \langle x+1| \otimes A_1+ |x+1\rangle \langle x| \otimes A_2+ |x\rangle \langle x| \otimes A_3,
\end{align}
where
\begin{align}
A_1=&R(\theta_1)P_0R(\theta_2)MR(\theta_2)P_0R(\theta_1),\\
A_2=&R(\theta_1)P_1R(\theta_2)MR(\theta_2)P_1R(\theta_1),\\
A_3=&R(\theta_1)P_1R(\theta_2)MR(\theta_2)P_0R(\theta_1)\nonumber\\&+R(\theta_1)P_0R(\theta_2)MR(\theta_2)P_1R(\theta_1),
\end{align}
with $P_0=|\uparrow\rangle\langle \uparrow|$ and $P_1=|\downarrow\rangle\langle \downarrow|$.

Writing the eigenstate as $|\Psi\rangle=\sum_{x,j} \beta_j^x|x\rangle\otimes|\psi_x\rangle$ and imposing the eigen equation $U|\Psi\rangle=\lambda|\Psi\rangle$, we have
\begin{align}
\det\left[\beta A_1+\beta^{-1}A_2+A_3-\lambda\right]=0.
\end{align}
While the equation above has two solutions $\beta_1$ and $\beta_2$, the open boundary condition imposes  one further condition $|\beta_1|=|\beta_2|$, from which we have
\begin{align}
|\beta|=\sqrt{\frac{\cosh\gamma\cos2\theta_2-\sinh\gamma}{\cosh\gamma\cos2\theta_2+\sinh\gamma}}.
\end{align}
Since $|\beta|$ is not a function of its phase, the GBZ of the system is always circular, with a radius $r=|\beta|$. In Fig.~\ref{fig:Sgbz}, we show the variation of the GBZ along the blue and red paths in the phase diagram (Fig.~\ref{fig:windingE}) of the main text. While the GBZ remains the same along the blue path, its radius changes along the red one. Incidentally, the radius approaches zero or infinity at the non-Bloch exceptional point [vertical green dashed lines in Fig.~\ref{fig:Sgbz}(a)] for our model.

\section{Projecting dynamics onto the GBZ}
In this section, we outline the projection of the time-dependent density matrix onto the GBZ.

Following Ref.~\cite{prr}, we first define the biorthogonal basis
$|k_{\beta,R(L)} \rangle=\frac{1}{\sqrt{N}}\sum_{x} r^{\pm x} e^{ik_\beta x}|x\rangle$ , where $N$ is the total number of unit cells and $r$ is the radius of GBZ. These basis states satisfy the biorthogonal and completeness relations: $\langle k_{\beta,L}|k'_{\beta,R} \rangle=\delta_{k_\beta,k_\beta'}$ and $\sum_{k_\beta}|k_{\beta,R}\rangle \langle k_{\beta,L}|=\one_{\text{w}}$, with $\one_{\text{w}}=\sum_x|x\rangle\langle x|$.
We then define the projection operator
$P_{k_\beta}=|k_{\beta,R}^f\rangle \langle k_{\beta,L}^f|\otimes \one_\sigma$, with $\one_\sigma=|\uparrow\rangle\langle \uparrow|+|\downarrow\rangle\langle\downarrow|$. This gives rise to a projected density matrix $\rho(k_\beta,t)=P_k |\psi(t)\rangle \langle \psi(t)|P^\dag_k$ in each generalized quasimomentum sector.

Dynamics of the projected density matrix is then captured by the dynamic spin texture
$\bm{n}(k_\beta,t)$, defined as~\cite{iS,XPNC}
\begin{align}
\bm{n}(k_\beta,t)=\frac{\text{Tr}\left[\rho(k_\beta,t)\eta\bm{\tau}\right]}{\text{Tr}\left[\rho(k_\beta,t)\eta\right]},\label{eq:nkS}
\end{align}
where $\bm{\tau} =(\tau_x, \tau_y, \tau_z)$, with $\tau_i=\sum_{\mu\nu=\pm}|\psi^f_{k_\beta,\mu}\rangle \sigma^{\mu\nu}_i \langle \chi^f_{k_\beta,\nu}|$ ($i=x, y, z$). Here, $\sigma^{\mu\nu}_i$ are the elements of Pauli matrices $\sigma_i$.
$|\psi^f_{k_\beta,\pm}\rangle$ ($\langle \chi^f_{k_\beta,\pm}|$) is the right (left) eigenstate of $H^f_{k_\beta}$, where
$H^f_{k_\beta}|\psi^f_{k_\beta,\pm}\rangle=E^f_{k_\beta,\pm}|\psi^f_{k_\beta,\pm}\rangle$ ($H^{f\dag}_{k_\beta}|\chi^f_{k_\beta,\pm}\rangle=E^{f\ast}_{k_\beta,\pm}|\chi^f_{k_\beta,\pm}\rangle$), $H^f_{k_\beta}=P_k H^f P_k$, the metric operator $\eta=\sum_{\mu}|\chi^f_{k_\beta,\mu}\rangle\langle \chi^f_{k_\beta,\mu}|$, and $\pm$ are the band indices. Here $\bm{n}(k_\beta,t)$ is a unit, real vector that can be visualized on the Bloch sphere.

As a example, consider the quasi-localized initial state used for our experiment: $|\Psi^i\rangle\propto|0,\uparrow\rangle+|0,\downarrow\rangle+2i/r_i|-1,\uparrow\rangle+1/r_i^2(-|-2,\uparrow\rangle+|-2,\downarrow\rangle)$, which is an eigenstate of the initial Floquet operator $U^i$ with $\theta^i_1=3\pi/4$ and $\theta^i_2=0.4\pi$. Here $r_i$ is the radius of the GBZ for $U^i$ under the open boundary condition.
Formally writing the initial state as
$|\Psi^i\rangle=1/\sqrt N\sum_{k_\beta} |\Psi^i_{k_\beta}\rangle$, where $|\Psi^i_{k_\beta}\rangle=|k^i_{\beta,R}\rangle\otimes|\psi^i_{k_\beta}\rangle$,
we evolve it under the Floquet operator $U^f=e^{-iH^f t}$. The time-evolved state is  $|\Psi(t)\rangle=e^{-iH^ft}|\Psi^i\rangle \equiv \sum_x |x\rangle \otimes |\psi _x (t)\rangle$. We then project it onto the GBZ of the final effective Hamiltonian $H^f$, using the relation
\begin{small}
\begin{align}
\rho(k_\beta,t)=&P_{k_\beta}\rho(t)P_{k_\beta}^\dagger\nonumber\\
=&\frac{1}{N}\sum_xr_f^{-x}e^{-ik_\beta x}|\psi_x(t)\rangle \sum_{x'} r_f^{-x'}e^{ik_\beta x'}\langle \psi_{x'}(t)|\nonumber\\
=&\frac{1}{2N} \sum_j\sum_{x,x'}r_f^{-(x+x')}e^{-ik_\beta(x-x')}\langle\psi_x(t)|\sigma_j|\psi_{x'}(t)\rangle \sigma_j.
\end{align}
\end{small}

The dynamic spin texture is then calculated from Eq.~(\ref{eq:nkS}) using the projected time-dependent density matrix $\rho(k_\beta,t)$. To project the density matrix onto the conventional BZ, one simply needs to replace the non-Bloch wave vector $k_\beta$ with $k$ and take $r=1$ in the definition of the biorthogonal basis states $|k_{\beta,R(L)} \rangle$, which then reduces to the Bloch states.

\begin{figure*}[htp]
\centering
\includegraphics[width=0.8\textwidth]{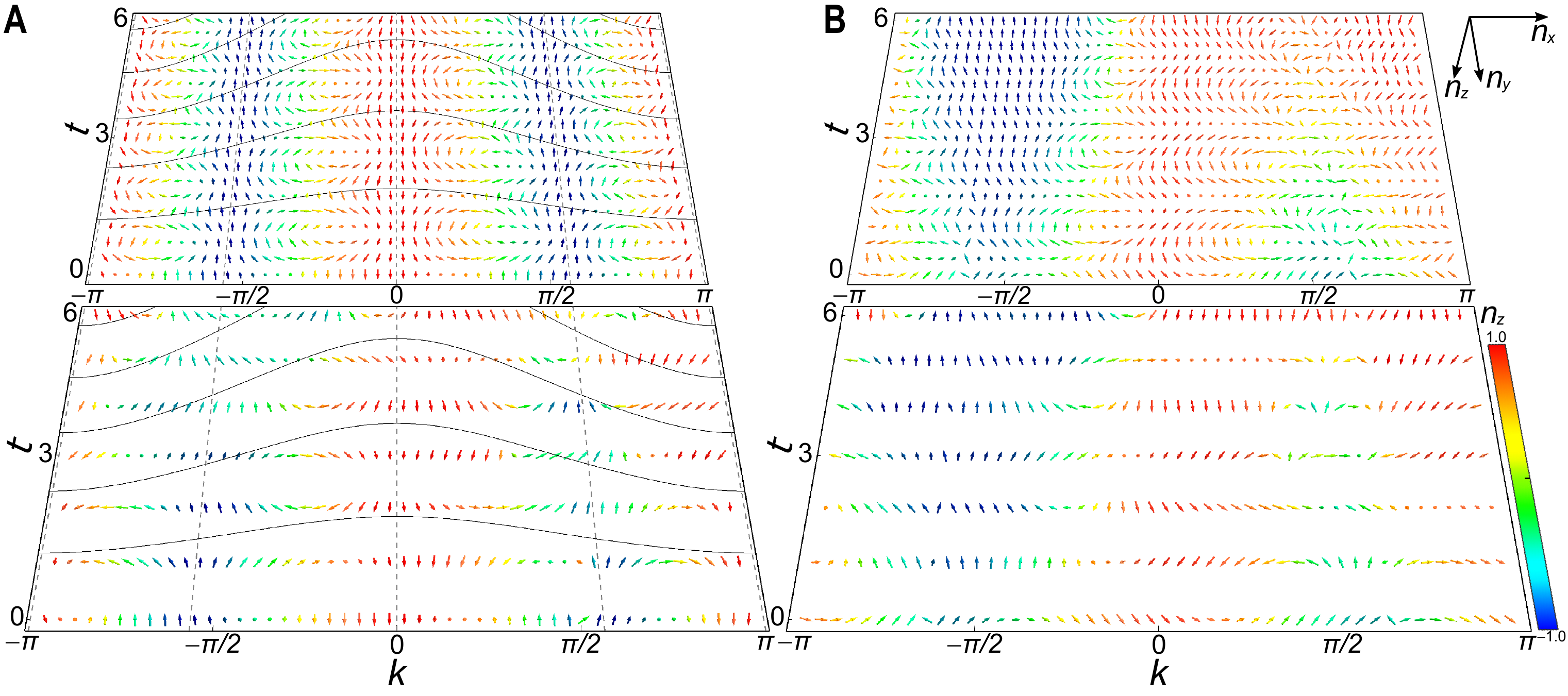}
\caption{(a) Dynamic spin textures $\bm{n}(k_{\beta},t)$ for a quench process between distinct non-Bloch topological phases with the same GBZ radius $r_i=r_f=1.41$. The initial Floquet operator is characterized by $\left(\theta_1=3\pi/4,\ \theta_2=0.4\pi\right)$ with $\nu^i_\beta=1$, and the final one characterized by $\left(\theta_1=0.33\pi,\ \theta_2=0.4\pi\right)$ $\nu^f_\beta=-1$. Compared to Fig.~\ref{fig:S1}, more fixed points exist together with the emergence of more complicated skyrmion structures, due to the larger difference between the initial and final non-Bloch winding numbers. (b) Dynamic spin texture $\bm{n}(k,t)$ of the same quench process as in (a), but projected into the BZ using the Bloch band theory.
Here $\nu^i=1$ and $\nu^f=-1$.
}
\label{fig:S2}
\end{figure*}

\begin{figure*}[htp]
\centering
\includegraphics[width=0.8\textwidth]{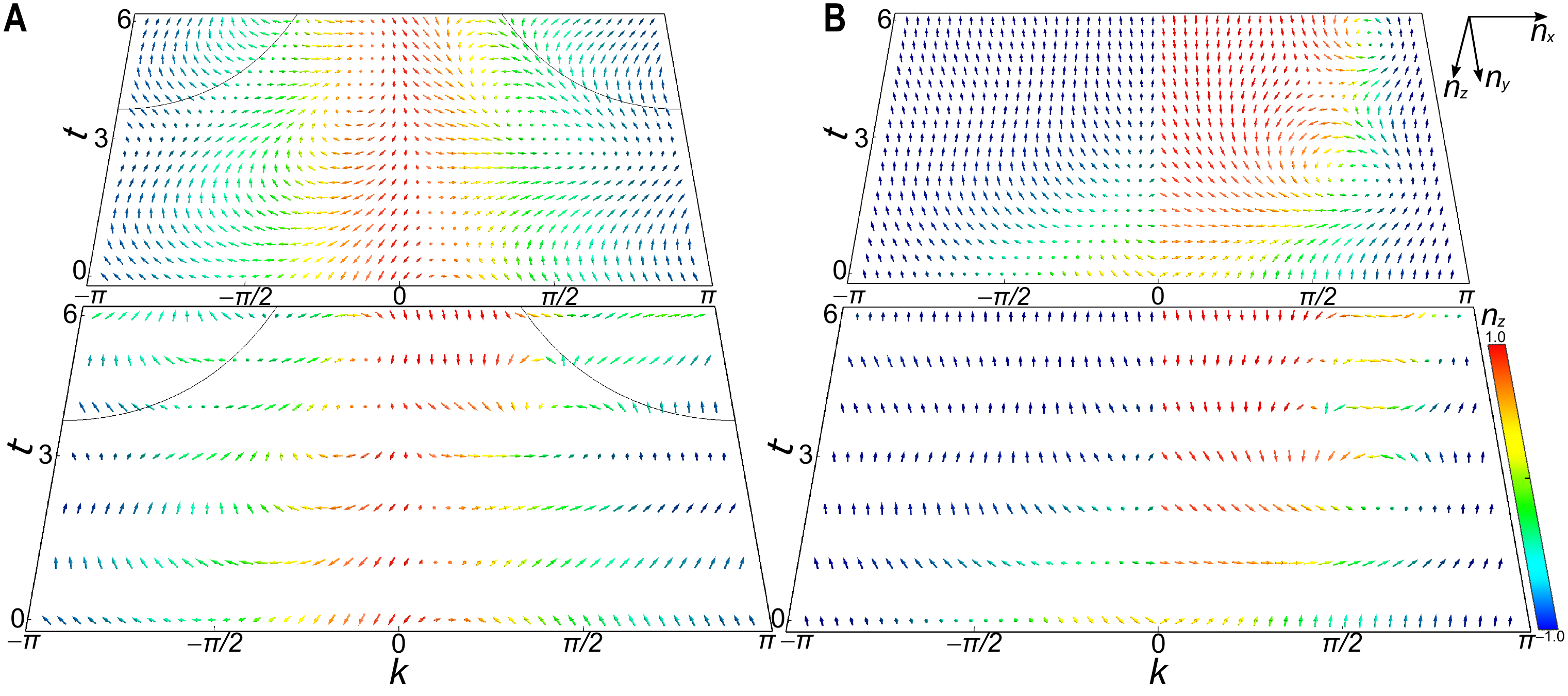}
\caption{(a) Dynamic spin textures $\bm{n}(k_{\beta},t)$ for a quench process between distinct non-Bloch topological phases with the GBZ radii $r_i=1.41$ and $r_f=2.23$. The initial Floquet operator is characterized by $\left(\theta_1=3\pi/4,\ \theta_2=0.4\pi\right)$ with $\nu^i_\beta=1$, and the final one characterized by $\left(\theta_1=0.666\pi,\ \theta_2=0.316\pi\right)$ with $\nu^f_\beta=0$. Despite the absence of fixed points as $r_i\neq r_f$, the global skyrmion structure persists.
(b) Dynamic spin texture $\bm{n}(k,t)$ of the same quench process as in (a), but projected into the BZ using the Bloch band theory. Here $\nu^i=1$ and $\nu^f=1/2$. Both $U^i$ and $U^f$ are in the non-Bloch exact PT phase.
}
\label{fig:S3}
\end{figure*}

\section{Biorthogonal local marker and chiral displacement}

In this section, we analysis how the time-averaged biorthogonal chiral displacement converges to the non-Bloch winding number. Invoking the biorthogonal local marker, our derivations here closely follow those in Refs.~\cite{sciencecd,ZCW}.

The biorthogonal local marker is defined as
\begin{equation}
\nu(x)=\frac{1}{4}\sum_s \langle {x},s| Q \Gamma \left[X,Q\right]|{x},s\rangle+h.c.,
\end{equation}
where $|{x},s\rangle$ labels the state on sublattice $s$ of the ${x}$th unit cell, $\Gamma=\one_\text{w} \otimes\sigma_x$ is the chiral symmetry operator, and $X$ is the unit-cell position operator. The biorthogonal projection operator $Q=P_{+}-P_{-}$, with $P_{\pm}= \sum_{n} | \psi_{n,\pm} \rangle \langle \chi_{n,\pm}|$. Here $|\psi_{n,\pm} \rangle$ is the $n$th right eigenstate of $U$, satisfying $U|\psi_{n,\pm} \rangle=\lambda_{n,\pm} |\psi_{n,\pm} \rangle$; and $ \langle\chi_{n,\pm} |$ is the $n$th left eigenstate, with $\langle \chi_{n,\pm}|U=\langle \chi_{n,\pm}|\lambda_{n,\pm} $. Due to the chiral symmetry of $U$, we have $|\psi_{n,\pm} \rangle=\Gamma|\psi_{n,\mp} \rangle$. It follows that
\begin{equation}
\begin{aligned}
\nu(0)=&\frac{1}{4}\sum_s \langle 0,s| Q \Gamma \left[X,Q\right]|0,s\rangle+h.c.\\
=&\sum_s \langle 0,s| P_- \Gamma XP_-|0,s\rangle+h.c.,\\
\end{aligned}
\end{equation}
where we have used $X|0,\uparrow \rangle=0$, $\left[X,\Gamma\right]=0$, and $Q={\bf 1}-2P_-$. In previous studies, it has been shown that the biorthogonal marker converges to the non-Bloch winding number under open boundaries~\cite{ZCW}.

We now turn to the biorthogonal chiral displacement defined in the main text, where
\begin{align}
C(t)=&\langle \chi (t)|\Gamma X|{\psi} (t)\rangle+h.c. \nonumber\\
=&\langle \chi (t)|(P_++P_-)\Gamma X(P_++P-)|\psi (t)\rangle +h.c.\nonumber\\
=&(\langle \chi (t)|P_+\Gamma XP_+|\psi (t)\rangle+\langle \chi (t)|P_-\Gamma XP_-|\psi (t)\rangle+\langle \chi (t)|P_+\Gamma XP_-|\psi (t)\rangle+\langle \chi (t)|P_-\Gamma XP_+|\psi (t)\rangle) +h.c.
\label{eq:SC}
\end{align}
We have numerically checked that the last two terms on the right-hand side of the expression above average out over sufficiently long time, and the first two terms are independent of time, provided: i) the system features completely real quasienergy spectra, i.e., it is in the non-Bloch exact PT phase; ii) the system size is sufficiently large. It follows that, to evaluate the time-averaged biorthogonal chiral displacement $\bar{C}$ for sufficiently long times, one simply needs to take $t=0$ in Eq.~(\ref{eq:SC}).
We therefore have
\begin{equation}
\begin{aligned}
\overline C =& \langle 0,\uparrow| P_{+} \Gamma X P_{+} |0,\uparrow \rangle + \langle 0,\uparrow| P_{-} \Gamma X P_{-} |0,\uparrow \rangle +h.c.\\
=&2\langle 0,\uparrow| P_{-}\Gamma X P_{-} |0 ,\uparrow\rangle +h.c.\\
=& \nu(0) .
\end{aligned}
\end{equation}
For the derivation, we have used $ \langle 0,\uparrow| P_{+} \Gamma X P_{+} |0,\uparrow \rangle =  \langle 0,\uparrow| P_{-} \Gamma X P_{-} |0,\uparrow \rangle$, and $ \langle 0,\uparrow| P_{-}\Gamma X P_{-} |0 ,\uparrow\rangle =\langle 0,\downarrow| P_{-}\Gamma X P_{-} |0 ,\downarrow\rangle$.

\begin{figure*}
\centering
\includegraphics[width=0.8\textwidth]{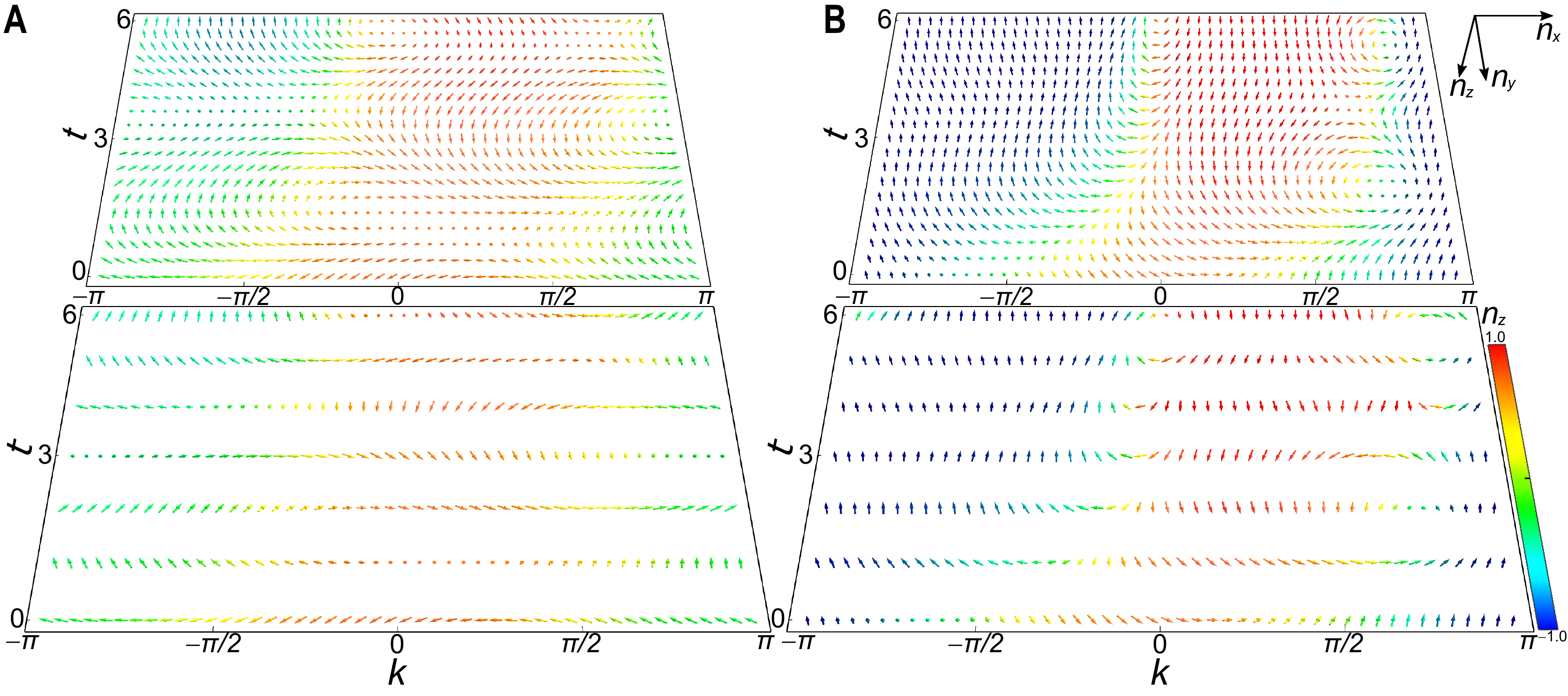}
\caption{(a) Dynamic spin textures $\bm{n}(k_{\beta},t)$ for a quench process between distinct non-Bloch topological phases, but with $U^f$ in the non-Bloch PT broken regime. The initial Floquet operator $U^i$ is characterized by $\left(\theta_1=3\pi/4,\ \theta_2=0.4\pi\right)$ with $\nu^i_\beta=1$, and the final Floquet operator $U^f$ is characterized by $\left(\theta_1=0.64\pi,\ \theta_2=0.29\pi\right)$ with $\nu^f_\beta=0$. The GBZ radii are $r_i=1.41$ and $r_f=5.19$.
(b) Dynamic spin texture $\bm{n}(k,t)$ of the same quench process as in (a), but projected into the BZ using the Bloch band theory. Here $\nu^i=1$ and $\nu^f=0$.
}
\label{fig:S4}
\end{figure*}

\section{Experimental data on skyrmions structures}

In this section, we present supporting experimental data on the dynamic spin textures for different quench processes. As discussed in Ref.~\cite{prr}, for a quench process between $U^i$ and $U^f$ with distinct non-Bloch winding numbers, fixed points of the dynamics exist in the GBZ, provided $U^i$ and $U^f$ have the same circular GBZ and are both in the non-Bloch exact PT phase. An exemplary dynamics at the fixed points is shown in Fig.~\ref{fig:fp}(a). The presence of fixed points enables the definition of dynamic Chern numbers on the submanifolds of generalized quasimomentum-time domain, which are essentially the skyrmions numbers of the emergent dynamic skyrmions. When $U^i$ and $U^f$ have different GBZ radii, fixed points can only exist in a perturbative sense. This is illustrated in Fig.~\ref{fig:fp}(b).
Dynamic skyrmions are still discernable and provide a practical detection scheme for the non-Bloch winding numbers, so long as the difference in the GBZ radius is not too large.

In Fig.~\ref{fig:spinT} of the main text, we compare the spin textures between the non-Bloch and Bloch quench descriptions, with $U^i$ and $U^f$ featuring distinct non-Bloch winding numbers. In Fig.~\ref{fig:S1}, we show the spin textures when $U^i$ and $U^f$ have the same non-Bloch winding numbers. As illustrated in Fig.~\ref{fig:S1}(a), skyrmions are absent in this case, while fixed points of the dynamics persist (dashed lines). Here the measured $\Delta \bar{n}_z$ is close to zero.
For the Bloch quench description, the spin dynamics are steady-state approaching (Fig.~\ref{fig:S1}(b)), due to the complexity of the quasienery spectra under the periodic boundary condition.

In Fig.~\ref{fig:S2}(a), we show the spin dynamics for $\nu^i_\beta=1$ and $\nu^f_\beta=-1$. While the measured $\Delta \bar{n}_z$ approaches zero in this case, a rich dynamic skyrmion structure emerges. A closer examination of the skyrmion structure shows that the vanishing $\Delta \bar{n}_z$ is due to the emergence of an additional fixed point between $k_\beta=0$ and $k_\beta=\pi$, which originates from the larger difference between $\nu^i_\beta$ and $\nu^f_\beta$. As fixed points demarcate different submanifolds in the generalized quasimomentum-time space on which dynamic skyrmions appear, mapping out the dynamic skyrmions structures is a more robust way to determine the non-Bloch topological invariants than measuring $\Delta \bar{n}_z$ alone.

While the quench processes above feature $U^i$ and $U^f$ with circular GBZ of the same radius, in Fig.~\ref{fig:S3}, we demonstrate the dynamic spin textures of a typical quench process between those with distinct radii. Compared to Fig.~\ref{fig:S1}, a main difference is that fixed points exist but in a perturbative sense. However, the global structure of dynamic skyrmions persists in the generalized quasimomentum-time domain, provided  $U^i$ and $U^f$ have distinct non-Bloch winding numbers.

Finally, to showcase the importance of the reality of quasienergy spectra for our detection scheme, we demonstrate the measured and simulated dynamic spin textures when $U^f$ is in the non-Bloch PT broken regime. As shown in Fig.~\ref{fig:S4}, dynamics in the generalized quasimomentum-time domain under the non-Bloch description (Fig.~\ref{fig:S4}(a)) is very similar to that under the Bloch description (Fig.~\ref{fig:S4}(b)), both featuring steady-state-approaching behavior.

\end{document}